# Investigating Yang-Mills theory and Confinement as a function of the spatial volume

A. González-Arroyo and P. Martínez,

*Departamento de Física Teórica C-XI, Universidad Autónoma de Madrid,*

*28049 Madrid, Spain.*

June 1995

**Abstract**

We study the volume dependence of electric flux energies for $SU(2)$ gauge theory with twisted boundary conditions. The curves interpolate smoothly between the perturbative semiclassical results and the Confinement regime. On the basis of our results, we propose that the Confinement property might be caused by a class of non-dilute multi-instanton configurations.

# 1 Introduction

Confinement has been the subject of active study since the early days of QCD. Wilson [1] clarified its meaning as a possible phase of a system of Yang-Mills fields, and put forward the relevant order parameter. 't Hooft added the characterization of other phases [2]. It was clear from the start that the occurence or not of Confinement was a complicated non-perturbative issue. Wilson's lattice formulation [1] casted the problem into that of a peculiar type of statistical mechanical system, opening the way to the powerful techniques developed in that field of research.

During the first years, most work on the subject was concentrated on either proving exactly the Confinement property, or in computing by numerical methods the value of the string tension $\sigma$ ( the constant whose non-zero value signals Confinement). The first goal has not been accomplished to date except in low dimensions, abelian theories [3] or in the presence of supersymmetry [4, 5]. On the numerical side, however, there is a clear signal of a string tension value close to the phenomenological-experimental one.

In the last years, nonetheless, more emphasis has been set on understanding how the Confinement regime arises. This amounts to having a description of the vacuum state of the theory and an identification of the relevant degrees of freedom that produce Confinement. The literature is full of proposals [6], mostly incompatible with each other. One of the most acclaimed ideas is the Dual Superconductor mechanism of 't Hooft and Mandelstam [7, 8]. Lately this idea has been translated and studied on the lattice [9, 10].

Our line of attack follows the work of Luscher [11] and collaborators who used the volume of space as a probe for the dynamics of Yang-Mills fields. As one moves from small to large volumes, one is moving, without a phase transition, from the region where Perturbation theory is a good approximation, to the one where neither the finiteness of space nor the boundary conditions play any role in local quantities. By observing how this process takes place we aim at a better understanding of the most relevant degrees of freedom in explaining the properties of Yang-Mills theory, most notably Confinement.

Our starting point differs from that of Luscher and Koller and van Baal

[12] in the use of twisted boundary conditions (t.b.c.). Although irrelevant for large volumes, the b.c.'s are crutial at small ones. Our study can be considered, hence, complementary to the one of the previous authors. We have, furthermore, reasons to believe that twisted b.c.'s are perhaps more efficient in showing up the most relevant configurations. Just on very general grounds, the twisted b.c.'s point towards disorder, which for a gauge system is Confinement. Actually, in the large N limit t.b.c.'s seem crutial in eliminating the volume dependence of the string tension [13]. To exemplify what we do have in mind, consider the rather trivial case of the one dimensional Ising model. We know that this model has no ordered phase. One can explain this fact in terms of the kink configurations ( a configuration with a single negative bond). The non-existence of a phase transition can be traced back to the non-zero probability to create a kink. Periodic b.c.'s imply an even number of kinks, while t.b.c.'s (antiperiodic ones) imply an odd one. The low temperature expansion in the first case starts with an ordered configuration. T.b.c.'s show nonetheless the right situation from the start and the leading order gives us a single kink state. This pattern is followed in other less trivial examples.

In this paper we are considering the case of SU(2) Yang-Mills theory with a spatial twist given by a magnetic flux vector $\vec{m} = (1, 1, 1)$ ( modulo 2 ). An additional bonus of the twisted situation is the discreteness of the space of classical vacua (modulo gauge transformations), which makes the weak coupling expansion free of the complications of the periodic case. This expansion, which is a good approximation for small spatial size, has been studied in Refs. [14] , [15], and [16]. The first two references deal with perturbation theory alone. In the third one, we studied the role played by configurations that tunnel between the classical vacua. That work was based on Monte Carlo simulations in a region of small sizes, where we tested that the semiclassical formuli were in agreement with the data, leading to the numerical determination of the scale factor, which is in principle, but not in practice, calculable analytically. Hence, at least in the quantities that were studied in Ref. [16] we have a fairly good analytical understanding of the low $l_s$ region.

The present paper is to be considered a continuation of Ref. [16] in which we explore the larger $l_s$ values and bridge the gap towards the Confinement regime. Some preliminary results were presented at the 1993 Lattice conference [17]. In the next Section we give the details of the numerical method employed to obtain our data. In the third one we discuss the extraction of the physical results from the data. In the fourth section we explore some ideas suggested by our data, and end up presenting an scenario for understanding the onset of the Confinement mechanism. In Section 5 a short summary of our results is presented.

## 2 The Data

In this section we present the technical details about the simulation. The data has been obtained on the transputer based machine RTN, located at the University of Zaragoza. Some results (those for smaller lattices) have been obtained from an 8-transputer board, similar to that contained in RTN, located at Universidad Autonoma. Details about RTN and the board can be found elsewhere [18]. The code, written in OCCAM2 language, consists on a heat bath Monte Carlo simulation for $SU(2)$ gauge fields on asymmetric lattices and twisted boundary conditions. Our lattices are symmetric in space, and the total volume is $N_s^3 N_t$. We have studied $N_s = 4, 6$ and 8 with $N_t = 128, 128$ and 64 respectively. We have been using Wilson action with a range of $\beta$ values going from 2.25 up to 2.6. The specific choices has been dictated by the need to cover a range of physical spatial lengths $l_s = N_s a$, in a region where scaling is seen to apply. We have tried to obtain equal values of $l_s$ for different values of $N_s$, in order to test scaling with our data. The twisted boundary conditions has been introduced in our data by flipping the sign of the contribution of all spatial plaquettes sitting at the corner of each plane $\mathcal{P}_{ij}(n_i = n_j = N_s)$.

All our simulations started with several tens of thousands of sweeps intended to bring our statistical sample to equilibrium. The starting configuration has varied between the cold one, a hot (random) one and a wrongly cold one corresponding to no twist. This variation enables us to test thermal-ization by comparing results with different initial conditions. In principle, the problem is more important for larger $\beta$ values, which were studied in our previous publication [16]. Summarising our tests, we can say that we have not found any signal of dependence on initial conditions in our results.

Measurements are taken every $N_{skip}$ M.C. sweeps. Typically $N_{skip}$ is chosen on the basis of the observed autocorrelation of different quantities. Local quantities are quickly uncorrelated, but some of our extended Polyakov loops are more strongly correlated. In any case the value of $N_{skip}$ which is finally selected is always chosen to be larger or equal to the observed exponential autocorrelation times for our observables. Tests have been made that the results do not change significantly by taking $N_{skip}$ larger than the one actually selected to present our results. Again in this case, the stronger autocorrelations occur for larger values of $N_s$ and $\beta$. The discussion presented in Ref [16] is certainly sufficient to sustain our confidence in our results. To conclude, we mention that the total number of M.C. iterations and measurements performed by our group is collected in Table 1, together with the remaining parameters employed in our simulation. Now we turn our attention to the actual measurements and observables.

We measure ground state energies in each electric flux sector by studying the time dependence of correlators of the appropiate operators, namely Polyakov loops $P$. These quantities act non-trivially with respect to the group of singular gauge transformations in space, or equivalently carry non-zero electric flux:

$$\vec{e} = \vec{\omega}(\gamma) \qquad (modulo\ 2). \tag{1}$$

The electric flux vector $\vec{e}$ [19, 20] is a 3-vector of integers modulo 2, while $\vec{\omega} \equiv (\omega_1, \omega_2, \omega_3)$ is the vector of winding numbers of the loop $\gamma$ around each of the 3 directions of space. In principle, one can choose any operator with the right quantum numbers to study these correlators. In practice, a good choice is required to get a reasonably good signal to background ratio. Fuzzying [21, 22] and Smearing [23, 24] are two techniques used in the literature to attain this goal. In our work we have used both techniques. The common denominator of both techniques is that, instead of employing straight Polyakov loops, one uses operators involving sums of many lattice paths with the right winding. This serves two purposes: it disminishes the fluctuations, due to uncorrelated link fluctuations, like $1/\sqrt{N_\gamma}$ ($N_\gamma$ is the number of paths involved), and, being a more extended operator, it is intended to increase the overlap with the ground state. Actually, use of these techniques is crucial to get reasonably small errors.

In our version of fuzzying, we perform a renormalization group transformation to a lattice of half the spatial extension. The new links being the normalised sum of all spatial length 2 and 4 paths between the same initial and final points. Straight Polyakov lines in the blocked lattice are what we call "block 1" operators. If we repeat the operation we get a lattice of one fourth the spatial extension and "block 2" operators. This can be done for $N_s = 4$ and 8. For $N_s = 6$, the second blocking is built of length 3 and 5 lines. Twisted Boundary conditions are taken into account by adding the different paths with appropiate signs. The sign is chosen so that all paths contribute equally in the large $\beta$ limit. All of the results that we will present involve *block-2* Polyakov loops $P^{b=2}$.

Our version of the other procedure, called Smearing, is simply a local cooling [25, 26] step, but for each time slice independently. This procedure has the advantage over the previous one that it can be iterated indefinitely. At each "smearing step", every link is replaced by the normalised sum of all spatial staples with the same initial and final points. Not all our simulations included smearing steps. Those that did, together with the number of these steps are shown in Table 1.

As an alternative to smearing, some of our simulations included several ordinary four-dimensional cooling steps. Computing straight line Polyakov loops on cooled configurations can be considered as simply using some new *cooled* operators on the original configuration. Nonetheless, the new operators are extended in time and the correlators of them might significantly change if the extension of these operators in time is comparable with their separation. One cooling step is, nevertheless, a local operation and hence, as a rule of thumb, we have limited our analysis to separations that are always larger than the number of cooling steps c.

In conclusion, every $N_{skip}$ Monte Carlo iterations we measured the fol-lowing quantities: $P_i^{b,s,c}(n)$ , $P_{ij}^{b,s,c}(n)$ , $P_{123}^{b,s,c}(n)$; where $b$, $s$ and $c$ are the number of applied blocking, smearing and cooling steps, n is a lattice point and $i$ runs over spatial directions. More precisely, we shall take $b = 2$ always, and cooling and smearing where never used at the same time due to memory limitations (See Table 1 for the characteristics of each simulation).

We then averaged every operator of the list over spatial points. Due to the boundary conditions, for the $|\vec{e}| = 1$ and 2 case, this is actually not projecting over an eigenstate of momentum. The Fourier descomposition of an operator $Q(\vec{n})$ is

$$Q(\vec{n}) = \frac{1}{N_s^3} \sum_{\vec{q}} e^{i\frac{2\pi}{N_s}\vec{q}\cdot\vec{n}} \ \hat{Q}_{\vec{q}} \ e^{i\frac{\pi}{N_s}\vec{s}\cdot\vec{n}} \tag{2}$$

where $\vec{s} = (\vec{e} \times \vec{m})$ ($\vec{e}$ and $\vec{m} = (1,1,1)$ are the electric and magnetic flux vectors), $\vec{q}$ is a 3-vector of integers ( modulo $N_s$) and $\vec{p} = \frac{2\pi}{N_s}(\vec{q} + \frac{\vec{s}}{2})$ is the momentum value. Hence, if we average over space, we get

$$\frac{1}{N_s^3} \sum_{\vec{n}} Q(\vec{n}) = \frac{1}{N_s^3} \sum_{\vec{q}} \hat{Q}_{\vec{q}} \prod_i \delta(s_i, q_i) \tag{3}$$

where $\delta(0,q) = \delta_{q,0}$ and

$$\delta(1,q) = \frac{i \ e^{i\frac{\pi}{N_s}(q+\frac{1}{2})}}{N_s \ sin(\frac{\pi}{N_s}(q+\frac{1}{2}))}. \tag{4}$$

For $s = 1$ all momenta appear in the sum, the smaller ones with higher coefficients. The only problem of using these operators instead of the pure eigenstates of momentum, is that the correlators may receive contributions from higher momentum states together with higher mass excitations. To ensure that this does not affect our results, we have also measured the operators corresponding to lowest momentum, at some values of $N_s$ and $\beta$. The results are consistent within errors, with no systematic deviation observed between using one type of operators or the other.

# 3  Analysis of the data

We have measured correlators of different Polyakov loops, leading to estimates of the minimum energies in all electric flux sectors. In the Confinement regime, one expects that the presence of a non-zero 't Hooft type of topological electric flux, leads to the formation of a string carrying this flux. This physical string must have the same winding number as the Polyakov line which creates the electric flux. The energy of such physical string is given by the string tension $\sigma$ times the length of the string. Actually, the minimal length turns out to be $l_s\,|\vec{e}| = l_s\,\sqrt{e_1^2 + e_2^2 + e_3^2}$, where $e_i \in \{0,1\}$ is the $i^{\text{th}}$ component of the electric flux vector and $l_s$ is the length of the torus. In conclusion, the minimum energies (relative to the $\vec{e} = \vec{0}$ vacuum) in each sector $E_{\vec{e}}$ should behave as

$$E_{\vec{e}} = |\vec{e}|l_s\sigma \ . \tag{5}$$

What we have actually done is to plot things in terms of the quantities

$$\Sigma_{|\vec{e}|^2} = \frac{E_{\vec{e}}}{|e|l_s}. \tag{6}$$

The Confinement prediction is that, for large $l_s$, $\Sigma_{|\vec{e}|^2}$ should tend to a constant, equal to the string tension $\sigma$, and independent on the value of $\vec{e}$. Actually, the last property (equallity of all $\Sigma_{|\vec{e}|^2}$) can be interpreted as a consequence of the recovery of rotational invariance.

In order to determine the energies, we measured correlations of operators with the appropriate quantum numbers $O(t)$ (the corresponding Polyakov loops)

$$C(t) = \frac{1}{N_t}\sum_{n=1}^{N_t}\langle O(n)\,O(n+t)\rangle \tag{7}$$

from which we extracted lattice energy estimates:

$$M(t+1) = -\ln\left(\frac{C(t+1)}{C(t)}\right) \tag{8}$$

The smaller $t$, the larger the systematic errors due to contamination with excited states, but usually the smaller the statistical errors. For self-adjoint operators one has that $M(t)$ decreases with $t$. A customary compromise is to give as estimate of the energies the value of $M(t)$ obtained for the smallest $t$ which is compatible with larger $t$'s within errors.

In our case, we have chosen $M(3)$ as an estimate for the energy. Actually, even $M(2)$ satisfies the afore-mentioned criterion in most cases; and even when it does not, there is no clear systematics, and deviations seem to be due to statistical fluctuations. Nonetheless, the errors for $M(2)$ are much smaller than those of $M(3)$, which makes the second value a safer choice. Additional confidence follows from the fact that we have used different operators with the same quantum numbers to measure the same mass, always with consistent results. The different operators are the ones associated with different smearing steps and both the momentum eigenstates, as the space averages, when available. When cooling was used, according to the reasoning of the last section, we have actually taken $M(3+c)$ as our estimate. Again compatibility within errors was attained. Our conclusion is that, either there is little contamination of higher states, or else the latter have energies that differ from the ground state ones by less than the errors.

In order to give the reader an idea of the quality of our data, we have collected in Table 2 a set of determinations for several operators and values of $N_s$, $\beta$ and time separations $t$. Errors have been estimated by partitioning. We show values of $\Sigma_i$ in physical units ( see below) for small ( 0.671 fm.), intermediate ( 0.910 fm.) and large volumes (1.233 fm.). The data at $\beta = 2.3$ and $N_s = 4$ represents two independent runs with up to 3 cooling steps, but differing by the use of momentum eigenstate operators ( MEO) or simple averages. The intermediate volume result given in Table 2 employed up to 9 smearing steps. Finally, the large volume one again consisted on two independent runs, one with cooling and the other with smearing. The $c = s = 0$ *Mixed* result combines the statistics of both runs. All our results are represented by converting energies into physical units by use of the empirical formula

$$a(\beta) = 400 \; exp(-\frac{log2}{0.205}\beta) \; fm. \tag{9}$$

which applies in our region of $\beta$ values according to the results of Ref. [27]. The absolute normalization is fixed by convention to $\sigma = 5 fm^{-2}$.

In Fig. 1, $\Sigma_{|\vec{e}|^2}$ is converted into physical units ($fm^{-2}$) and plotted as a function of $l_s$. We have added the results of Ref. [16], which affect only $\vec{e} = (1,1,1)$ and $l_s < 0.7\ fm$. The plotted values follow from the $M(3+c)$ estimates of the energies. The best choice for the number of coolings $c$ or the number of smearing steps $s$, is selected. Notice that in some cases data from different lattice sizes give very similar values of $l_s$. Our data shows that scaling is satisfied to within a few percent.

At low values of $l_s$ the predictions of the semiclassical approximation are also shown. Both $\Sigma_1$ and $\Sigma_2$ receive a contribution from perturbation theory, going like $\frac{1}{l_s^2}$. The coefficients were computed in Ref. [15]. $\Sigma_3$ is zero to all orders in perturbation theory, and becomes non-zero as a result of tunneling mediated by the $Q = \frac{1}{2}$ instanton [28]. In Ref. [16] we showed that indeed the data follow the predictions of the semiclassical dilute gas approximation. This determines the $l_s$ dependence of $\Sigma_3$ up to a multiplicative prefactor $A_3(N_s)$, which, in the absence of an analytical computation, has been fitted to our data. In the case of $\Sigma_1$ and $\Sigma_2$, we also allowed for a non-perturbative piece of the same shape but with a free prefactor.

Hence, the curves shown in Fig. 1 correspond to the functions:

$$\Sigma_i(N_s, \beta) = \left(A_i(N_s)\beta^2 e^{-\frac{\beta}{4}S(N_s)} + B_i(N_s) + \frac{C_i(N_s)}{\beta}\right)\frac{1}{l_s^2}, \qquad (10)$$

where $S(N_s)$ is the lattice action of the $Q = \frac{1}{2}$ instanton, $B_i(N_s)$ and $C_i(N_s)$ ($i = 1, 2$) are given in Ref. [15], and $A_i(N_s)$ is fitted to our data. Actually, for $|\vec{e}|^2 \neq 3$, the dominant term is the one in the middle, where $B_1 = B_2$ is equal to 4.14, 4.30 and 4.36 for $N_s = 4$, 6 and 8 respectively. The next perturbative piece has the right tendency for the splitting between $\Sigma_1$ and $\Sigma_2$ shown by the data, but its magnitude is very small (1 %). Finally, the non-perturbative piece gives contributions ranging from 3 to 10 % as one moves from $l_s = 0.4$ to $0.7 fm$. Due to the size of the errors, setting $A_2 = A_3 = 0$ still gives rise to a reasonably good fit, and the curves sit in between the $\Sigma_1$ and $\Sigma_2$ data. The many curves seen in Fig 1. come from the different values of $N_s$. Their closeness is again a manifestation of scaling for our data.

In summary, we can say that the semiclassical predictions describe our data fairly well, confirming that we have a reasonably good understanding of the dynamics occuring for volumes smaller than $0.7 fm$, which is the border of the dilute gas approximation region. On the opposite side, for large $l_s$, Confinement tells us that the different $\Sigma$'s should merge into a unique horizontal line of height $5 fm^{-2}$. Indeed our data show this tendency, and furthermore the interpolation between both regions is fairly smooth. This behaviour is the main result of our data, and in the next section we will try to understand how this could take place.

# 4   An scenario for Confinement

The purpose of this section is to analyse the implications of our Monte Carlo results on the nature of the $SU(2)$ gauge theory vacuum and Confinement. We expect that these findings could be translated to $SU(3)$ theory and $QCD$. The most important conclusions from our data are two. First, it determines the scale at which the finite volume theory approaches the infinite volume one to within a 20% accuracy. Our quantities $\Sigma_{|\vec{e}|^2}$ actually point towards $1.0 - 1.2$ $fm$ as the appropiate scale. Furthermore, precisely at this scale rotational invariance is recovered with the same level of accuracy. Second, the approach to the infinite volume value is smooth. There seems to be no scale where a qualitative change of behaviour is observed. This is reinforced if we take into account that also the glueball spectrum is qualitatively alike for large and small volumes [29]. This is in contrast with the situation for periodic boundary conditions [30].

Hence, it is natural to expect as well a smooth transition between the configurations which dominate the path integral for different values of $l_s$. Of course, it is logically possible that a qualitative change in the structure of vacuum does not lead to drastic changes in the values of $\Sigma$. We nevertheless would like to explore the implications of our more natural hypothesis. Indeed, we have a very good understanding of the structure of the vacuum for $l_s < 0.7$ $fm$. It consist on a gas of dilute twisted periodic instantons [28] modulated by the gaussian (and higher) perturbative corrections. This is seen with textbook clarity for very small $l_s$, and describes perfectly well our data for $l_s \lesssim 0.7$ $fm$ as seen in Fig. 1 and Ref [16]. Hence, we are lead to conjecture that the descendants of the twisted periodic instantons do populate the Yang-Mills vacuum. This is very shocking at first sight, and it is the purpose of this section, to clarify how this can happen and to extract some physical consequences from this fact. We will end up proposing an scenario which has several appealing features. One of these, is the possibility of explaining both the value of the string tension and the topological susceptibility in terms of the same classical configurations. Even more, it is possible that chiral symmetry breaking also originates from them [31].

Before proceeding to describe what can happen for $l_s \simeq 1.2\ fm$, let us consider again the small $l_s$ region. The number of $Q = \frac{1}{2}$ instantons per unit time $n_I$, keeps increasing with $l_s$. Indeed, the time extent of each instanton is $l_s$, so that the dilute gas approximation breaks down when

$$n_I \simeq \frac{1}{l_s} \ . \tag{11}$$

Given the relationship between $\Sigma_3$ and $n_I$, we have that this occurs at $l_s = \overline{l_s}$ such that

$$\Sigma_3(\overline{l_s}) \simeq \frac{2}{\sqrt{3}} \frac{1}{\overline{l_s}^2} \ , \tag{12}$$

and this can be seen to hold for $\overline{l_s} \simeq 0.7\ fm$. Hence, beyond this point the instantons are very close to each other and one has to take into account their interaction. Within the dilute gas approximation there is a relationship between the topological susceptibility and $\Sigma_3$ which is as follows

$$\chi = \frac{\sqrt{3}}{8 l_s^2} \Sigma_3 \ . \tag{13}$$

Thus, at the border of the dilute gas approximation, one has $\chi \sim 1\ (fm)^{-4}$, which is in agreement with what other groups have measured in their Monte Carlo simulations [32].

What happens beyond $0.7 fm$ ? There are several possibilities. $Q = \frac{1}{2}$ instantons might fuse and make ordinary $Q = 1$ instantons. However, the latter are ineffective in producing Confinement and hence, if this was the case we should observe a decrease in the string tension precursor $\Sigma_3$. It is very unnatural to think that some other mechanism could come in precisely at that point and keep the things slowly varying. There is indeed something else which we know must happen: some other configurations can start to play a role in the path integral. In particular one can consider the descendants of our twisted periodic instanton. These configurations are periodic in space but with periods which are fractions of $l_s$. Let us note this period by $d$. We then have that $l_s/d$ is an integer . Notice that these configurations are special types of multi-instanton configurations on the torus ( or in $\infty$ space). They exist both for twisted or periodic boundary conditions, with only an overall constraint on their number distinguising different b.c.'s.

How special are these configurations? If we have one such periodic configuration with $N$ lumps, it has a total topological charge of $Q = N/2$. According to the index theorem [33] there are $8Q = 4N$ linearly independent deformations of these solutions into other multi-instanton solutions. Only 4 of the total $4N$ maintain the periodic arrangement. Notice that the number of degrees of freedom of the multi-instanton space equals that of a 4-dimensional gas of particles. We propose to view the multi-instanton space precisely as a gas of $Q = \frac{1}{2}$ lumps, where one needs only to give the centers of these lumps to specify the configuration uniquely. These lumps behave as elastic balls or balloons, whose size and shape adjusts appropiately given their location. The lumps are not isolated or dilute ( in all directions); they cannot be, since their size grows with separation. It is rather a liquid than a gas. This description is obviously possible and valid in the neighbourhood of the periodic configurations. For large deformations it must fail. When two lumps are pushed into each other, they fuse into a single $Q = 1$ lump, and their 8 parameter space now gives the position, size and orientation of the lump (as for ordinary instantons). By generating configurations on the lattice, we have actually observed how there is a continuous path from the two $Q = \frac{1}{2}$ lump structures to the $Q = 1$ ones [34].

Summarising, we are proposing to consider a finite measure part of the multi-instanton space, which can be described as a liquid of $Q = \frac{1}{2}$ lumps. We will label such space by the mean-separation $d$ among lumps, which is limited by the finite spatial size and the boundary conditions. We will deduce the properties of these configurations from those of the periodic ones, of which they are deformations. An analytic and numerical understanding of this parametrization of multi-instanton space and its properties is a purely classical problem. By now only numerical evidence in favour has accumulated, but the problem is currently under study.

How do these configurations become relevant at larger volumes? To study this point, we saturate the path integral with this kind of configurations. The structures are naturally 3 dimensional for small volumes, and so we will make the counting in a way which privileges time. We have

$$Z = \sum_{n(d)} \frac{e^{-n(d)F(d,l_s)}}{\prod n(d)!} \tag{14}$$

where $n(d)$ is the number of space-periodic configurations with period $d$ (a fraction of $l_s$). The relative weight of one such configuration is $e^{-F(d,l_s)}$. If we make the simplifying hypothesis (which we will comment upon later) that

$$e^{-F(d,l_s)} = \left(e^{-F(d,d)}\right)^{(\frac{l_s}{d})^3} , \tag{15}$$

this relative weight is a function of $exp\{-F(d,d)\}$ which can be extracted from our data in the region where the dilute gas approximation applies:

$$l_s E_{\vec{e}=(1,1,1)} = 2e^{-F(l_s,l_s)} \sim (l_s/\overline{l_s})^\alpha \tag{16}$$

with $\alpha \sim 2$ and $\overline{l_s} \simeq 0.6 - 0.7$ $fm$. This formula can be assumed to be valid, by extrapolation, for $l_s$ slightly larger than $\overline{l_s}$ as well. In the thermodynamic limit ($l_s \longrightarrow \infty$ with d fixed), the relative weight peaks at $d = \overline{d}$ given by the maximum of the function

$$G(d) = \left(e^{-F(d,d)}\right)^{1/d^3} \tag{17}$$

Indeed, the previous function has a maximum which is independent of $\alpha$, given by $\overline{d} = \overline{l_s} e^{1/3} \gtrsim \overline{l_s}$. Although the precise relation between $\overline{d}$ and $\overline{l_s}$ depends on the details of the computation, the existence of a maximum and the fact that $\overline{d} \gtrsim l_s$, can be easily understood. For $d < \overline{l_s}$ it is indeed more probable to have the perturbative vacuum than to have an instanton of size $d$ in one cell of size $d^4$. Actually, since these instantons are periodic in space, the larger the volume the more suppressed such a configuration is. The tendency changes when $d > \overline{l_s}$. Then the more instantons the more probable. However, the probability of one instanton increases with $d$, but the number of instantons that fit into a fixed spatial volume decreases with $d$. Hence there is a maximum within this region.

Although our approximations are presumably fairly crude, the conclusion is quite robust to modifications: In the the thermodynamic limit ($l_s \longrightarrow \infty$), only one of the family of periodic configurations with period $d$ dominates the path integral, namely that with $d = \overline{d}$ the maximum of $G(d)$. This leads to the following description of what could happen as the volume of space increases from 0 to $\infty$. For $l_s \lesssim \overline{d} \sim \overline{l_s}$ the dominating configuration of instantons has $d = l_s$ and we have a dilute gas of these configurations. Beyond $l_s \sim \overline{d}$ we start having configurations with periods $d = l_s/n$ (n integer). The most probable one actually corresponds to $d = \overline{d}$ independently of $l_s$. This explains how one is finally going to obtain results which are volume independent for large volumes. The results will also be independent on the boundary conditions imposed in space: periodic or twisted. Actually, boundary conditions only fix the evenness or oddness of the number of periodic instantons contained in space, which is irrelevant for large numbers.

Now we have to face the meaning and justification of formula (15). Actually, $F(d, l_s)$ gives the free energy relative to the vacuum of the periodic instanton configuration. In the semiclassical approximation one has:

$$e^{-F(d,l_s)} = \mathcal{A}(d, l_s) e^{-\frac{4\pi^2}{g^2}(l_s/d)^3} \qquad (18)$$

where the exponential term is $exp\{-\frac{S_{cl}}{g^2}\}$ and $S_{cl}$ is the classical action of the configuration. In our case this action is $4\pi^2$ for each cell of size $d^3$, in agreement with (15). Hence, we simply need to prove and analogous property for the prefactor $\mathcal{A}(d, l_s)$. This quantity incorporates the integration over gaussian fluctuations around the classical configuration and the integration over the collective coordinates of the zero-modes. For gaussian fluctuations of high frequency ($\gg 1/d$) it is reasonable to expect that the fluctuations around each configuration are insensitive to the environment in which they are placed. This translates again into a dependence of type (15). Deviations from this formula are to expected for frequencies $\in [\frac{1}{l_s}, \frac{1}{d}]$, and our hypothesis amounts to assuming that their effect does not disturb expression (15) too much. Concerning zero-modes, notice that to compute $\mathcal{A}(d,d)$ one has to consider the only four zero-modes associated to space-time translation of the configuration. To justify (15) we should have 4 zero-modes for each cell of size $d^3$, which is precisely what we expect from the index theorem. Notice, that irrespective of $l_s$, one expects that each collective coordinate associated to a zero-mode is of range $d$. We stress that, since we are integrating over the zero-mode space, we are actually considering not-only spatially periodic configurations, but also the other non-periodic configurations: The whole set of positions of the gas of lumps in our scenario. Our main assumption in this case is that gaussian fluctuations around these non-periodic configurations are similar to those for periodic ones. In conclusion, expression (15) is the most reasonable approximation to be made in the absence of an exact expression. We expect that this assumption gives qualitatively correct results although quantitative corrections are likely.

The previous considerations imply that the main free parameter in our instanton liquid model of the vacuum, the mean separation d, is fixed dynamically in the average to a value close to $0.7 fm$. We will see that such a class of configurations gives rise to properties of the vacuum which are in reasonable qualitative agreement with the known properties of the SU(2) gauge vacuum. First we will consider the topological susceptibility $\chi$ and next the string tension.

Assuming as usual that there are equal probabilities for $Q = \frac{1}{2}$ instantons and anti-instantons, the topological susceptibility turns out to be equal to

$$\chi = \frac{4}{d^4} \simeq 1 fm^{-4} . \qquad (19)$$

This value is in agreement with the results of Monte Carlo measurements of other authors [32]. Actually, since our instantons are non-dilute, there is no reason why the probability of instantons and anti-instantons should be independent. There seems to be some evidence in the direction of a higher probability of alike charges to cluster [35], but this might be also due to the distortion caused by cooling. In any case, at our qualitative level, a reasonable amount of clustering can be compensated with a small change in $\overline{d}$.

Finally, we should touch upon a crucial point. Does the proposed liquid give rise to a non-vanishing string tension? Although it might be possible to show such a behaviour by numerical methods, we do not yet have results in this direction. Here, we will try to make a crude analytical approximation which is expected to give the right order of magnitude. Again our considerations will make use of our knowledge of the periodic $Q = \frac{1}{2}$ configuration in a 4 dimensional twisted box of size $\overline{d}$: our prototype of an individual lump [36]. If we measure the square Wilson loop around the boundaries of the box centered at the lump, we find that if the position of the loop in the transverse directions coincides with the center of the loop, then the value of the Wilson loop is $-I$. On the other hand if the Wilson loop lies in between two lumps, the result is $I$. At intermediate positions the loop gives an element of the group interpolating between both, but not situated along a single $U(1)$ subgroup of SU(2). Since computing Wilson loops in classical configurations is notably difficult, we will make use of the following customary abelianizing approximation. We will assume that the loop is only either $-I$ or $I$. The first case occuring when the loop passes within a distance $x\overline{d}$ from the center of the lump. Otherwise the result is $I$. Typically this approximation amounts to replacing a continuous function going from -1 to 1, by a step function. The parameter $x$ should lie between 0 and $\frac{1}{2}$. Now, if we have a large Wilson loop, its value in a given configuration is equal to $(-1)^M$, where M is the number of lumps which are within a distance $x\overline{d}$ from the minimal surface of the loop. In our gas picture, M has a Poissonian-binomial distribution. The value one gets for the string tension is:

$$\sigma = \frac{2}{\overline{d}^2} g(x) \simeq 4g(x) fm^{-2} \qquad (20)$$

where $g(x)$ depends on the details of the distribution: $4x^2$ for Poissonian, $-ln(1-8x^2)/2$ for a binomial one. Notice that for g(x) of the order of 1, one gets a value of the string tension which is close to the one measured for large lattices. Since the string tension depends on $\overline{d}^{-2}$, a relatively small change in $\overline{d}$ can adjust for corrections to the approximation.

# 5 Conclusions

In this paper we have presented a follow up of the evolution of the value of electrix flux ground state energies as a function of the spatial volume. Our data nicely show how and where the predictions of Confinement are obtained for large lattices. No intermediate scales are appreciated in our data and at sizes of $l_s \simeq 1 - 1.5 fm$ both rotational invariance and Confinement predictions are well reproduced. Since an analytic understanding of the low $l_s$ region is known [16], we are led to speculate on the way that the corresponding structure of the vacuum smoothly evolves into the confining vacuum. An appealing scenario arises in which an important fraction of the space of multi-instanton configurations can be described as a liquid of (non-isolated) $Q = \frac{1}{2}$ instantons. From results of our previous work [16], we argue that the average density of lumps in the vacuum is dynamically fixed to be $(0.7 fm)^{-4}$ approximately. With standard assumptions, we show that this gas could lead to values of the topological susceptibility $\chi$, and of the string tension $\sigma$ which are close to the ones known to hold for SU(2) Yang-Mills theory.

Our scenario has in some respects striking similarities with the one proposed by Callan, Dashen and Gross [37]. First of all, the main constituents of our vacuum have $\frac{1}{2}$ topological charge, and can be combined in pairs and smoothly deformed to make ordinary instantons. These properties are shared by the corresponding objects in the case of Ref. [37], which are called merons [38]. Analogously, the mechanism by means of which merons produce Confinement is virtually the same that we have employed in computing the string tension for our liquid. Nonetheless, there are also important differences between our $Q = \frac{1}{2}$ instantons and merons. In our case, the configurations are smooth everywhere, while merons are singular solutions. It might be possible nonetheless, that one could establish some connection between our configurations and merons.

Unfortunately, unlike the meron case, there is no analytic formula for the $Q = \frac{1}{2}$ instantons, although they are quite well known numerically. This problem difficults progress on the analytic side. It is nevertheless possible to study these problems numerically on the lattice. The $Q = \frac{1}{2}$ lumps are seen to be present in our Monte Carlo generated configurations [39]. We are currently investigating the structure of the cooled SU(2) gauge vacuum in order to prove or disprove this or other models of Confinement. Some of our results are strikingly consistent with our model [40].

# Table captions

1. Table 1:

   We show the total list of simulations that we have performed, indicating the value of $\beta$ and $N_s$, the initial configuration, the number of measurements, the number of Monte Carlo sweeps between two sucessive measurements $N_{skip}$ and the number of thermalization sweeps $N_0$. We indicate how many cooling steps and smearing steps, if any, we have performed and whether we measure the three fluxes or only the one with $|\vec{e}|^2 = 3$.

2. Table 2: We give the values of $\Sigma_{|\vec{e}|^2}$ and errors for some of our simulation points ($N_s$ and $\beta$). The columns labelled $t_n$ indicate that the result is extracted from correlations at distances $c + n$ and $c + n + 1$, where c is the number of coolings. For each simulation, different estimates of $\Sigma$ from different operators is shown (s stands for smearing step).

# Figure captions

1. Figure 1:

   The values of $\Sigma_{|\vec{e}|^2}$ (defined in Eq. (6)) are plotted as a function of $l_s$. The different symbols used for different values of $N_s$ are shown in the plot. The semiclassical curves are described in the text and from top to bottom correspond to $\Sigma_1$ $N_s = 6$ and 4, $\Sigma_2$ $N_s = 6$ and 4 and $\Sigma_3$ $N_s = 4$, 6 and 8.

Table 1:

| $N_S$ | $\beta$ | Genesis | Meas. | $N_{skip}$ | $N_0$ | Type | Flux |
|---|---|---|---|---|---|---|---|
| 8 | 2.325 | cold | 568 | 500 | 40000 | s=0,3,6,9 | 1,2,3 |
| 8 | 2.325 | cold | 1336 | 250 | 324000 | c=0,1,2,3 | 1,2,3 |
| 8 | 2.360 | cold | 750 | 500 | 40000 | s=0,3,6,9 | 1,2,3 |
| 8 | 2.385 | hot | 310 | 500 | 40000 | s=0,3,6,9 | 1,2,3 |
| 8 | 2.415 | cold | 300 | 500 | 40000 | s=0,3,6,9 | 1,2,3 |
| 6 | 2.330 | cold | 644 | 200 | 40000 | s=0,3,6,9 | 1,2,3 |
| 6 | 2.369 | cold | 1211 | 100 | 56000 | s=0,3,6,9 | 1,2,3 |
| 6 | 2.395 | hot | 740 | 200 | 40000 | s=0,3,6,9 | 1,2,3 |
| 6 | 2.425 | cold | 1127 | 100 | 40000 | c=0,1,2,3 | 3 |
| 6 | 2.460 | cold | 620 | 250 | 40000 | c=0,1,2,3 | 1,2,3 |
| 6 | 2.483 | hot | 335 | 500 | 42500 | s=0,3,6,9 | 1,2,3 |
| 4 | 2.300 | hot | 818 | 100 | 40000 | c=0,1,2,3 | 1,2,3 |
| 4 | 2.300 | hot | 541 | 100 | 121800 | c=0,1,2,3 | 1,2,3 |
| 4 | 2.300 | cold | 3920 | 50 | 20000 | c=0,1,2 | 3 |
| 4 | 2.340 | hot | 808 | 100 | 70000 | s=0,3,6,9 | 1,2,3 |
| 4 | 2.340 | cold | 3920 | 50 | 20000 | c=0,1,2 | 3 |
| 4 | 2.360 | cold | 686 | 100 | 120000 | s=0,3,6,9 | 1,2,3 |
| 4 | 2.380 | hot | 1692 | 50 | 300000 | c=0,1,2,3 | 1,2,3 |
| 4 | 2.380 | hot | 642 | 50 | 384600 | c=0,1,2,3 | 1,2,3 |
| 4 | 2.400 | hot | 1692 | 100 | 40000 | c=0,1,2,3 | 1,2,3 |
| 4 | 2.425 | hot | 484 | 100 | 40000 | c=0,1,2,3 | 1,2,3 |
| 4 | 2.440 | hot | 319 | 250 | 20000 | c=0,1,2,3 | 1,2,3 |
| 4 | 2.475 | hot | 219 | 500 | 40000 | s=0,3,6,9 | 1,2,3 |

Table 2:

| $N_S$ | $\beta$ | Type | $|\vec{e}|^2 = 1$ | $|\vec{e}|^2 = 1$ | $|\vec{e}|^2 = 2$ | $|\vec{e}|^2 = 2$ | $|\vec{e}|^2 = 3$ | $|\vec{e}|^2 = 3$ |
|---|---|---|---|---|---|---|---|---|
|   | $(l_s)$ |   | $t_2$ | $t_3$ | $t_2$ | $t_3$ | $t_2$ | $t_3$ |
| 8 | 2.325 | s=0 | 5.20(13) | 4.80(31) | 5.84(29) | 4.9(1.5) | 4.77(61) | – |
| 8 | (1.233) | s=3 | 5.27(16) | 4.83(43) | 5.18(28) | 5.1(1.3) | 5.10(61) | 13(11) |
| 8 |   | s=6 | 5.42(18) | 4.59(42) | 5.20(32) | 5.2(1.9) | 5.04(67) | – |
| 8 |   | s=9 | 5.47(21) | 4.66(50) | 5.36(37) | 5.2(1.8) | 5.10(70) | – |
| 8 | 2.325 | c=0 | 5.029(73) | 4.62(22) | 5.30(19) | 4.28(55) | 4.43(33) | 3.9(2.0) |
| 8 | (1.233) | c=1 | 4.83(12) | 4.75(39) | 4.59(27) | 3.24(96) | 3.61(46) | 3.3(1.8) |
| 8 |   | c=2 | 4.79(26) | 4.8(1.1) | 3.83(74) | 2.9(1.7) | 2.8(1.1) | 3.5(1.8) |
| 8 |   | c=3 | 4.85(78) | – | 2.7(1.5) | – | 3.8(2.3) | – |
| 8 | MIXED | c=0 | 5.098(63) | 4.75(17) | 5.36(13) | 4.46(44) | 4.52(21) | 5.01(30) |
| 8 | 2.415 | s=0 | 6.36(15) | 6.41(28) | 6.64(20) | 6.40(53) | 4.18(20) | 3.69(48) |
| 8 | (0.910) | s=3 | 6.07(14) | 6.64(37) | 6.48(17) | 5.63(45) | 3.95(17) | 3.44(36) |
| 8 |   | s=6 | 6.08(17) | 6.62(40) | 6.36(18) | 5.75(49) | 3.69(20) | 3.56(28) |
| 8 |   | s=9 | 6.20(18) | 6.62(44) | 6.28(23) | 5.74(61) | 3.57(14) | 3.73(34) |
| 4 | 2.300 | c=0 | 9.25(18) | 9.90(58) | 8.40(25) | 7.85(62) | 2.837(51) | 2.821(77) |
| 4 | (0.671) | c=1 | 8.88(40) | 8.26(88) | 7.93(50) | 7.3(1.3) | 2.791(72) | 2.88(12) |
| 4 |   | c=2 | 8.35(71) | 8.7(1.6) | 7.2(1.1) | 4.6(2.2) | 2.87(10) | 2.99(26) |
| 4 |   | c=3 | 8.6(1.3) | 8.4(3.1) | 4.4(1.9) | – | 2.93(26) | 2.96(44) |
| 4 | 2.300 | c=0 | 9.03(27) | 8.58(62) | 7.92(31) | 7.9(1.1) | 2.847(62) | 2.667(77) |
| 4 | MEO | c=1 | 8.80(44) | 9.5(1.2) | 8.10(75) | 7.2(2.5) | 2.765(67) | 2.724(98) |
| 4 |   | c=2 | 9.15(97) | 11.9(3.1) | 6.9(1.6) | 4.7(5.7) | 2.780(92) | 2.82(18) |
| 4 |   | c=3 | 10.7(2.0) | 13.3(8.9) | 6.9(7.5) | – | 2.81(18) | 2.87(36) |

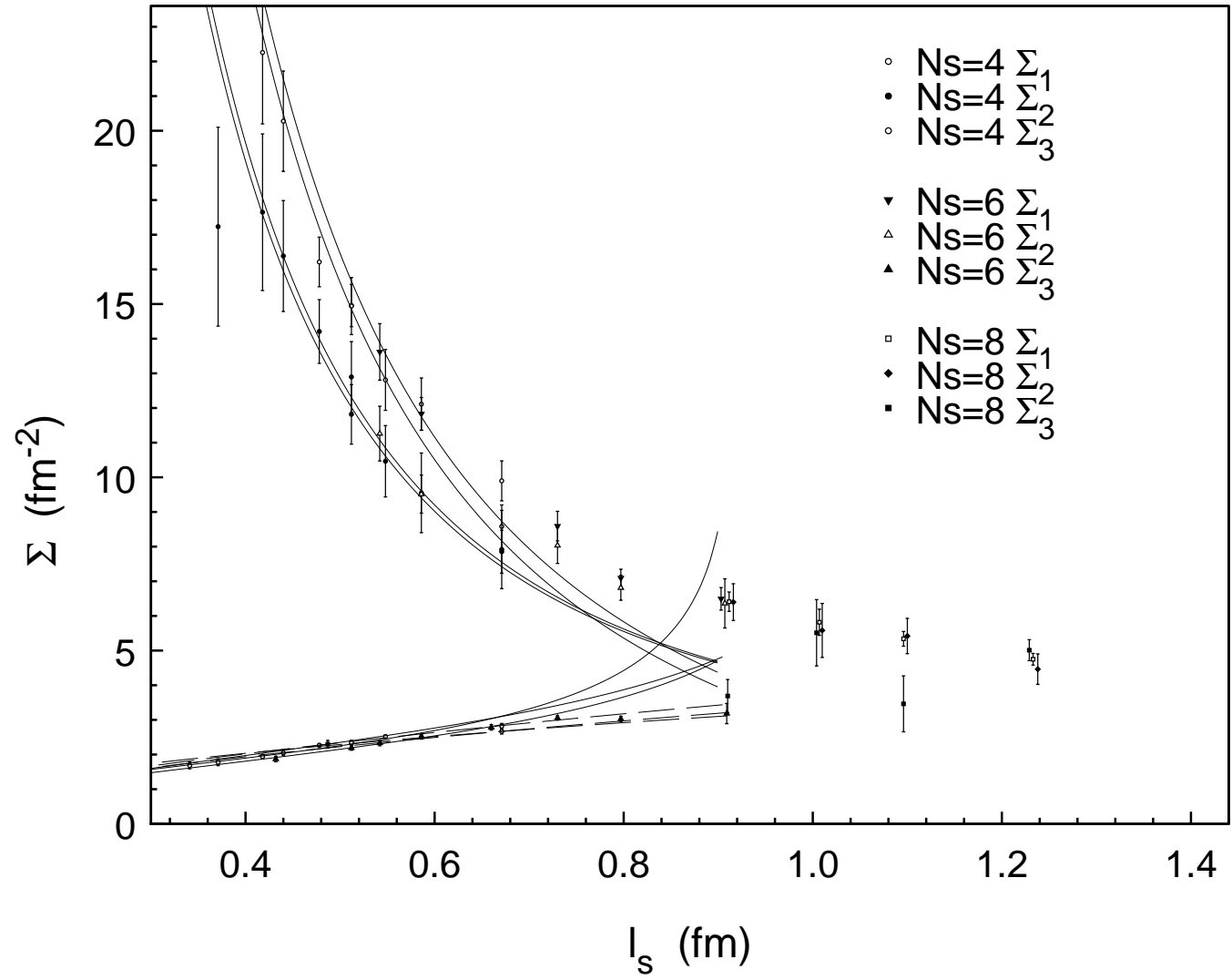

Fig. 1